\documentclass[aps,twocolumn,showpacs, showkeys,nofootinbib,floatfix]{revtex4-1}

\usepackage{amsmath}
\usepackage{amsfonts}
\usepackage{txfonts}
\usepackage{graphicx}
\usepackage[colorlinks=true, linkcolor=red, citecolor=blue]{hyperref}
\usepackage{ulem}

\begin{document}

\title{Distinguish three time-dependent dark energy models using statefinder pairs with error bars and Bayesian evidence}

\author{Hongchao Zhang}
\author{En-Kun Li} 
\author{Lixin Xu}
\email{lxxu@dlut.edu.cn}

\affiliation{Institute of Theoretical Physics, School of Physics and Optoelectronic Technology, \\ Dalian University of Technology, Dalian, 116024, People's Republic of China}

\begin{abstract}
	In this work, two completely different approaches, statefinder with error bars and Bayesian evidence, are used to distinguish and judge three time-dependent dark energy models. The parameters constrain for the three dark energy models are given using the current cosmic observational data sets : $Planck$ 2015, SNIa, BAO and OHD. Using the statefinder pairs with error bars, we find that the error region of the dark energy model, whose equation of state parameter is given by $w(a) = w_0+w_3\frac{1-a}{a^2+(1-a)^2}$, is relativity compact than the other two models during the all the evolving history. Meanwhile, the Bayesian evidence also provide that this model is significantly better than the other two models and the other two models are inconclusive. Then, there are reasons for believing that this model is a preferential candidate in dark energy investigation rather than the other two.
\end{abstract}

\maketitle

\section{Introduction}

Type Ia supernova (SNIa) observations \cite{ref:1-1,ref:1-2}, cosmic microwave background (CMB) anisotropy measurement from Wilkinson Microwave Anisotropy Probe (WMAP) \cite{ref:2-3}, and large scale structure from the Sloan Digital Sky Survey (SDSS) \cite{ref:2-4} show that our Universe is undergoing an accelerating expansion phase.
These observations reveals that about $70\%$ of cosmic components should be responsible for the last-time acceleration.
In order to explain the mechanism behind this phenomenon, theorists introduced an exotic energy component with negative pressure, called dark energy.
For recent reviews about dark energy, please see \cite{ref:2-5,ref:2-6,ref:2-7,ref:2-8,ref:2-9,ref:2-10,ref:2-11}.

Based on this opinion, various of dark energy models have been proposed.
Among all these models, $\Lambda$CDM model is the simplest and most attractive one at first \cite{ref:2-5}.
In standard $\Lambda$CDM cosmology, the late-time cosmic acceleration is provided by a cosmological constant satisfying an equation of state (EOS) $w \equiv P_{de}/\rho_{de} = -1$.
The latest released $Planck$ 2015 data \cite{ref:4-1} shows that Planck spectra at high multipoles $(l \gtrsim 40)$ are extremely well described by the standard spatially-flat six-parameter $\Lambda$CDM cosmology with a power-law spectrum of adiabatic scalar perturbations.
The results favour the base $\Lambda$CDM cosmology seemingly.
Moreover, by combining $Planck$ TT$+$lowP$+$lensing data with other astrophysical data, including JLA supernovae, the EOS of dark energy is constrained to $w = -1.006 \pm 0.045$ \cite{ref:4-1}.
This suggests that the observed Universe could be well described by $\Lambda$CDM model.

However, there are two thorny issues in $\Lambda$CDM cosmology, 
i.e. the ``cosmic coincidence problem'' and the ``fine tuning problem''. And the energy density of cosmological constant is $120$ orders of magnitude smaller than expected from quantum field theory \cite{ref:dark energy}.
Consequently people proposed a lot of alternative dark energy models, for related articles, please see \cite{ref:3-6,ref:3-15,ref:3-16,ref:3-22,ref:3-24,ref:3-27,ref:3-28,ref:3-31,ref:3-43,ref:3-46,ref:3-49,ref:3-52,ref:3-60,ref:3-64,ref:3-66,ref:3-70}.
So many alternatives, it becomes a crucial task to put forward an efficacious mechanism so that one can distinguish evolving dark energy models from a cosmological constant. Two completely different approaches are chosen to achieve this goal, that is, statefinder and Bayesian evidence.
The former is based on the dynamic evolution phase of specially cosmological objects, and the latter compare models from a statistical viewpoint.

In 2003, Sahni et al. \cite{ref:3-71,ref:3-72} proposed a geometrical diagnostic pair for dark energy, i.e. statefinder, which is constructed from the scale factor $a(t)$ and its derivatives up to the third order.
The statefinder pair \{r,s\} is defined as follows
\begin{equation}\label{1-1}
  r = \frac{\dddot{a}}{aH^2},~~~~ s = \frac{r-1}{3(q-1/2)}.
\end{equation}
Since the statefinder was put forward, researchers have used it to discriminate a great deal of dark energy models \cite{ref:3-73,ref:3-74,ref:3-75,ref:3-76,ref:3-77,ref:3-78,ref:3-80,ref:3-81,ref:3-82,ref:3-83,ref:3-84,ref:3-85,ref:3-87,ref:3-88,ref:3-89,ref:3-90}.
Recently, Arabsalmani et al. \cite{ref:Arabsalmani} generalized the original statefinder to statefinder hierarchy $S_n, S_n^{(1)}, S_n^{(2)}$, which are constructed by higher derivatives of scale factor, then Wang and Meng \cite{ref:3} utilized them to distinguish four time-dependent dark energy (TDDE) \cite{ref:3-32,ref:3-33,ref:3-34,ref:3-35,ref:3-36,ref:3-37,ref:3-38,ref:3-39} scenarios and the $\Lambda$CDM scenario from each other.

Statefinder gives a trajectory in phase diagram for one dark energy model.
As will be shown in subsequent section, these diagnostics can be expressed in terms of some fundamental quantities, the EOS of dark energy $w$, the fractional density of dark energy $\Omega_{de}$ (or baryons $\Omega_{m}$), and the scale factor $a(t)$.
In \cite{ref:3}, authors constrained TDDE models by using SNIa, baryonic acoustic oscillations (BAO), observational Hubble parameter (OHD) data sets as well as single data point from the newest event GW150914, they obtained best fitting values of model parameters \{$\Omega_{m0}$, $w_{0}$, $w_{i}$\} in TDDE models.
However, they dropped error bars unwarily.
It is possible for those diagnostics that may be very sensitive with these model parameters.
The trajectories should have error bars at $1\sigma$ or more higher confidence levels.
It is meaningless when there is only best fitting values without error bars.
Most of authors focused on these diagnostics only in the theoretical perspective without observational data, or someone used observational data to constrain their models, but no one considered the errors of the diagnostics that carried from the original variables.
Hence, we will constrain three TDDE models by using more observational data sets, and analyze their diagnostics in terms of the new values of model parameters with error bars via the error propagation equation. This is the first time to introduce the error propagation equation into the model diagnostic mechanism. In this way, the trajectories in phase diagram will be replaced by regions. It is worth to expect that more information about the difference of models would be revealed through this new idea.

Another effective and different approach to compare models is model selection in Bayesian methods.
Model selection quantifies how well the data conform to the overall predictions of a model, which depends on model dimensionality and model priors.
Bayesian inference consists of parameter estimation and model comparison.
Parameter estimation is generally performed by using Markov Chain Monte Carlo (MCMC) methods which is well known for us.
In order to perform model comparison, one must calculate the Bayesian evidence \cite{ref:bayesian book1}, which is a high-dimensional integration of the likelihood over the prior density.
An efficacious methodology for evaluating this integration is provided by nested sampling \cite{ref:nest sampling1}.
Several algorithms have been developed based on nested sampling, where MultiNest \cite{ref:multi-nest1,ref:multi-nest2} and CosmoNest \cite{ref:cosmo-nest} are the most widely used two.
However, they struggles with high-dimensional parameter spaces, and is unable to take advantage of this separation of speeds.
To address these issues, W.J. Handley et al. \cite{ref:ploychord1,ref:ploychord2} developed a new generation of algorithm, PolyChord \cite{ref:ploychord3}, which provide a means to sample high-dimensional spaces across a hierarchy of parameter speeds.

This paper is organized as follows.
In Section \ref{sec:models}, firstly, we recast the statefinder into new expressions which modified by the error propagation equation and did some essential review about three TDDE models at the same time.
Then a general overview of Bayesian inference is given.
In Section \ref{sec:constrain}, we constrain the models parameters and estimate the Bayesian evidence for three models.
Then in Section \ref{sec:analysis}, we discriminate and analyse three TDDE models by using the statefinder and the Bayesian evidence.
A summary is presented in the last section.

\section{Models and methodology} \label{sec:models}

\subsection{Statefinder hierarchy with error bars for three TDDE models} \label{sub:A}

Usually, cosmologists regard dark energy as a kind of ideal fluid, thus its physical properties can be characterized by EOS $w$, and fractional density $\Omega_{de}$ on background level.
Generally speaking, they are all functions of cosmic time $t$ or scale factor $a$.
By solving the Friedmann equation and the conservation equation, one can get the relation between the geometrical variable $H$ and the dynamical quantities
\begin{eqnarray}\label{2-1}
  E^2(a) \equiv && \frac{H^2(a)}{H^2_0} \nonumber \\
	=&& \Omega_{m0}a^{-3}+(1-\Omega_{m0})\exp \left[-3\int_1^a \frac{1+w(\tilde{a})}{\tilde{a}}d\tilde{a} \right],
\end{eqnarray}
where $E(a)$ is the dimensionless Hubble parameter.
Because of both energy densities of the baryons and the cold dark matter (CDM) evolve  proportional to $a^{-3}$, we express their density at present period together in terms of $\Omega_{m0}$.
Then $\Omega_{de0} \equiv 1-\Omega_{m0}$ can be read immediately.
The radiations have been neglected since we focus on the last-time acceleration.
Another important geometrical variable, the deceleration parameter $q$, can be expressed by those two physical quantities as
\begin{equation}\label{2-2}
  q \equiv -a\ddot{a}/\dot{a}^2 = \frac{1}{2} (1+3w(a)\Omega_{de}(a)),
\end{equation}
where
\begin{equation}\label{2-3}
  \Omega_{de}(a) = \frac{E^2(a)-\Omega_{m0}a^{-3}}{E^2(a)}
\end{equation}
is the relation of the fractional density of dark energy evolving over the scale factor $a$.
By the definition of the statefinder (\ref{1-1}), $r$, $s$ can be obtained from the derivative of deceleration parameter,
\begin{eqnarray}
  r &=& q(1+2q)-a\frac{dq}{da},\label{2-4} \\
  s &=& \frac{r-1}{3(q-1/2)}. \label{2-5}
\end{eqnarray}

However, to discriminate dark energy models better, Arabsalmani et al. proposed statefinder hierarchy as a new generation of diagnostic, which contains higher derivatives of the scale factor.
We define a dimensionless quantity as
\begin{eqnarray}\label{2-6}
  &A_n \equiv \frac{a^{(n)}}{aH^n},
\end{eqnarray}
where $a^{(n)}=d^na/dt^n$.
We follow the authors of \cite{ref:3} and define statefinder hierarchy $S_n$ as
\begin{eqnarray}
  &S_2 = A_2+\frac{3}{2}\Omega_{m}(a), \label{2-7} \\
  &S_3 = A_3, \label{2-8} \\
  &S_4 = A_4+\frac{3^2}{2}\Omega_{m}(a), \label{2-9} \\
  &S_5 = A_5-3\Omega_{m}(a)-\frac{3^2}{2}\Omega_{m}^2(a),~~~.... \label{2-10}
\end{eqnarray}
This statefinder hierarchy $S_n$ can be improved sequentially to another version $S^{(1)}_n$ to reply more complicated models,
\begin{eqnarray}
  &S^{(1)}_3 = S_3, \label{2-11} \\
  &S^{(1)}_4 = A_4+3(1+q),\label{2-12} \\
  &S^{(1)}_5 = A_5-2(4+3q)(1+q),~~~.... \label{2-13}
\end{eqnarray}
It is easy to verify that both hierarchies have the same property for the fiducial cosmology, that is
\begin{equation}\label{2-14}
  S^{(1)}_n\mid_{\Lambda CDM}=S_n\mid_{\Lambda CDM}=1.
\end{equation}
Therefore, for $\Lambda$CDM cosmology, the pair of statefinder have an apparent expression $\{S_n,S^{(1)}_n\}\mid_{\Lambda CDM}=(1,1)$ on the phase diagram. The second member of statefinder hierarchy can be constructed from $S^{(1)}_n$ in the following way \cite{ref:Arabsalmani},
\begin{equation}\label{2-15}
  S^{(2)}_n = \frac{S^{(1)}_n-1}{3(q-\frac{1}{2})}.
\end{equation}

Recently, in a series of articles, researchers studied various dark energy models by using those statefinder hierarchies. However, we inspect those statefinder hierarchies from another perspective, i.e. the error bars. We take two pairs of them, $\{S^{(1)}_3, S^{(1)}_4\}$ and $\{S^{(1)}_5, S^{(2)}_5\}$, as examples to illustrate the influence of the error bars on the results of distinguishing dark energy models. The concrete formulas of statefinder in terms of the cosmological parameters can be written as
\begin{eqnarray}
  &S^{(1)}_3 = A_3 = r, \label{2-16} \\
  &A_4 = r-3(q+1)r+a\frac{dr}{da}, \label{2-17} \\
  &S^{(1)}_4 = r-3(q+1)r+a\frac{dr}{da}+3(1+q), \label{2-18}\\
  &A_5 = A_4-4(q+1)A_4+a\frac{dA_4}{da}, \label{2-19}\\
  &S^{(1)}_5 = A_5-2(4+3q)(1+q), \label{2-20}\\
  &S^{(2)}_5 = \frac{S^{(1)}_5-1}{3(q-\frac{1}{2})}. \label{2-21}
\end{eqnarray}
Those formulas are all functions of $\Omega_{m0}$, $w(a)$ and $a$. In this paper, we parameterize EOS $w(a)$ in the following three time-dependent ways,
\begin{eqnarray}
  &w(a) = w_0+w_1\ln a, \label{2-23} \\
  &w(a) = w_0+w_2(1-a), \label{2-24}  \\
  &w(a) = w_0+w_3\frac{1-a}{a^2+(1-a)^2}, \label{2-25}
\end{eqnarray}
where $w_0$ denotes the present ($a=1$) value of $w(a)$. (\ref{2-23}) (\ref{2-24}) (\ref{2-25}) correspond to models 1, 2, 3 respectively. $w_0$, $w_i$ and $\Omega_{m0}$ are all free parameters for each models, thus they should be estimated by observational data sets. By using the CosmoChord in Section \ref{sec:constrain}, these parameters are coming with error bars. The errors would propagate from these parameters to the statefinder via the error propagation equation.

Generally speaking in statistics, people usually use a $n$ dimension vector $\mathbf{X}= (X_1,X_2,\cdot\cdot\cdot,X_n)^T$ to express $n$ random variables which can be obtained directly. And $\mathbf{\mu} =(\mu_1,\mu_2,\cdot\cdot\cdot,\mu_n)^T\equiv E(\mathbf{X})$ is the expected value of vector $\mathbf{X}$. For certain function $f$ of the $n$ random variables, the square error $\sigma^2_f$ of $f$ can be given by the following formula
\begin{eqnarray}\label{2-26}
  \sigma^2_f &\simeq&  \sum_{i=1}^{n}\sum_{j=1}^{n} (\frac{\partial f}{\partial X_i} \frac{\partial f}{\partial X_j})_{\mathbf{X}=\mathbf{\mu}} Cov(X_i,X_j) \nonumber \\
   &=& \sum_{i=1}^{n}(\frac{\partial f}{\partial X_i})^2_{\mathbf{X}=\mathbf{\mu}} Cov(X_i,X_i) \nonumber \\
  && +2\sum_{i<j,j=2}^{n} (\frac{\partial f}{\partial X_i} \frac{\partial f}{\partial X_j})_{\mathbf{X}=\mathbf{\mu}} Cov(X_i,X_j),
\end{eqnarray}
where $Cov(X_i,X_j) \equiv Cov(X_j,X_i)$ is the covariance matrix of $n$ random variables. Then $f$ with error bar should be rewritten as $f \pm \sigma_f$.

In this work, we consider a three dimension vector of random variances $(\Omega_{m0},w_0,w_i)$, and the corresponding covariance matrix can be generated during MCMC. Since the statefinder is time-dependent, while the error propagation equation is time-independent, thus the uncertainties can be translate to all the evolving history, i.e. arbitrary redshifts (even negative redshifts). When adding error bars in the phase diagram, it should be regions rather than trajectories. The positional relationships of these regions would be multifarious, and more information about dark energy models can be unscramble than before.

\subsection{Bayesian model comparison} \label{sub:B}

Usually, Bayesian inference consists of parameter estimation and model comparison. The former is used to constrain the parameters space of a single model, and can be implemented by MCMC, while Bayesian model comparison can help us to pick the most suitable model with current observational dataset among several models. The key of this process is the Bayes' theorem.

Given some datasets $\mathcal{D}$ and one model $\mathcal{M}$ with parameters $\theta$, one can estimate the posterior distribution on $\theta$ by using the Bayes' theorem
\begin{equation}\label{2-32}
  p(\theta|\mathcal{D},\mathcal{M}) = \frac{\mathcal{L}(\mathcal{D}|\theta,\mathcal{M})\pi(\theta|\mathcal{M})}{\mathcal{Z}(\mathcal{D}|\mathcal{M})},
\end{equation}
where $p(\theta|\mathcal{D},\mathcal{M})$ is the posterior probability distribution of the parameters, $\mathcal{L}(\mathcal{D}|\theta,\mathcal{M})$ is the likelihood, $\pi(\theta|\mathcal{M})$ is the prior, and $\mathcal{Z}(\mathcal{D}|\mathcal{M})$ is the Bayesian evidence. Since $p(\theta|\mathcal{D},\mathcal{M})$ is a probability distribution function for $\theta$, it has to be normalized to unity
\begin{equation}\label{2-33}
 1\equiv \int p(\theta|\mathcal{D},\mathcal{M}) d\theta =\frac{\int \mathcal{L}(\mathcal{D}|\theta,\mathcal{M})\pi(\theta|\mathcal{M})d\theta}{\mathcal{Z}(\mathcal{D}|\mathcal{M})},
\end{equation}
and therefore
\begin{equation}\label{2-34}
  \mathcal{Z}(\mathcal{D}|\mathcal{M})=\int \mathcal{L}(\mathcal{D}|\theta,\mathcal{M})\pi(\theta|\mathcal{M})d\theta.
\end{equation}
The function $\mathcal{Z}$ does not depend on the parameters $\theta$ and therefore it is of no help in estimating the parameters. From the viewpoint of posterior it is just a normalization factor.

To compare two models $\mathcal{M}_1$ and $\mathcal{M}_2$ based on the same observed dataset $\mathcal{D}$, one can calculate the ratio of their posterior \cite{ref:bayes-factor}
\begin{equation}\label{2-35}
  R=\frac{P(\mathcal{M}_1|\mathcal{D})}{P(\mathcal{M}_2|\mathcal{D})}=\frac{P(\mathcal{D}|\mathcal{M}_1)P(\mathcal{M}_1)}{P(\mathcal{D}|\mathcal{M}_2)P(\mathcal{M}_2)}=\frac{\mathcal{Z}_1}{\mathcal{Z}_2}\frac{P(\mathcal{M}_1)}{P(\mathcal{M}_2)}.
\end{equation}
Generally, one assumes that $P(\mathcal{M}_1)=P(\mathcal{M}_2)$ since there is not a prior reason for preferring one model over the other. It can be seen from equation (\ref{2-35}) that the Bayesian evidence plays a central role in model selection. Harold Jeffrey proposed a scale (TABLE \ref{tab:jeffrey}) in his book \cite{ref:jeffreys} which can be used as a criterion to compare models in terms of the logarithmic Bayesian evidence ratio.

Estimating the multi-dimensional integral in equation (\ref{2-34}) is a big challenge for computations. Nested sampling is a Monte Carlo technique to estimate the Bayesian evidence by transforming the multi-dimensional integral into a one-dimensional integral. We use a new generation of algorithm named PolyChord as a plug-in components in CosmoMC \cite{ref:2-36} to implement the nested sampling, termed CosmoChord. So that the parameters space and the Bayesian evidence can be obtained at the same time.

\begingroup
\begin{center}
\begin{table}
\renewcommand{\arraystretch}{1.8}
\begin{tabular}{ccc}
\hline\hline
$\triangle \ln \mathcal{Z} < 1.0$& ~ & Inconclusive \\
$1.0<\triangle \ln \mathcal{Z} < 2.5$& ~ & Significant \\
$2.5<\triangle \ln \mathcal{Z} < 5.0$& ~ & Strong evidence \\
$5.0<\triangle \ln \mathcal{Z} $& ~ &Dicisive \\
\hline\hline
\end{tabular}
\caption{Harold Jeffrey's scale for the interpretation of Bayesian evidence model comparison, where $\Delta\ln\mathcal{Z}= \ln(\mathcal{Z}_1/\mathcal{Z}_2)$. }\label{tab:jeffrey}
\end{table}
\end{center}
\endgroup

\section{The parameters constrain and Bayesian evidence} \label{sec:constrain}

Two kinds of diagnostic for three TDDE models have been analyzed in previous section. A global fitting for those models by using the cosmic observational data sets from $Planck$ 2015 \cite{ref:4-1} low-$l$, TT, EE, TE, lensing, joint SNIa \cite{ref:2-41,ref:2-42,ref:2-43}, BAO \cite{ref:2-38,ref:2-39,ref:2-40,ref:3-97}, OHD \cite{ref:3-98,ref:3-99} would be performed. The total likelihood $\chi^2$ can be constructed as
\begin{equation}
	\chi^2=\chi_{Planck}^2+\chi_{SNIa}^2+\chi_{BAO}^2+\chi_{OHD}^2,
	\label{3-1}
\end{equation}
where the right four terms denote the contributions from $Planck$ 2015, SNIa, BAO, OHD dataset respectively. The parameter spaces for the models considered in this paper is given by
\begin{equation}\label{3-2}
  P_i = \{\Omega_{m0}, w_0, w_i\},
\end{equation}
where $i=1,2,3$ corresponding to three TDDE models.

\begin{figure}
\begin{center}
  \includegraphics[scale=0.35]{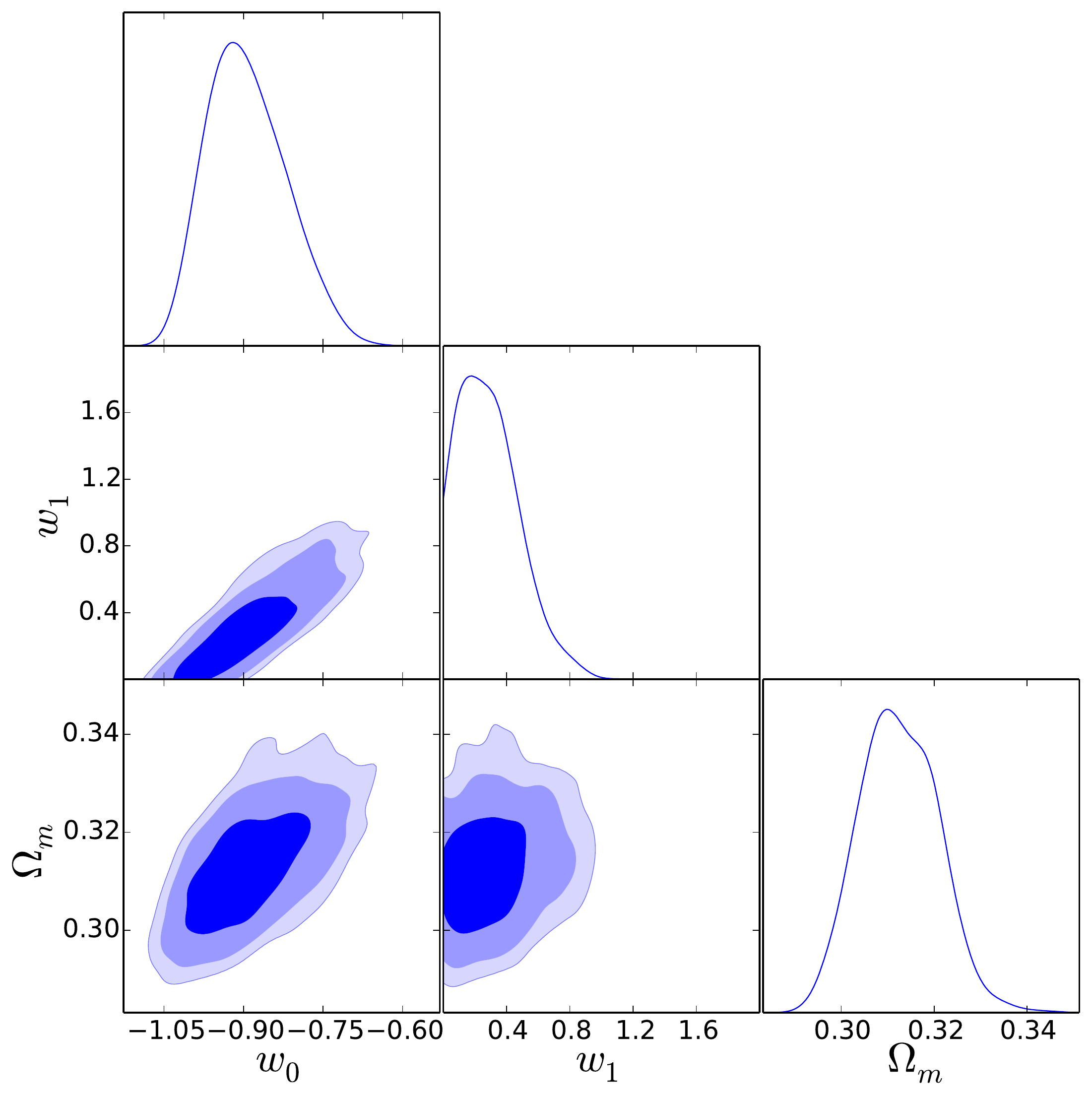}
\caption{1D marginalized distributions on individual parameters and 2D contours with 68\% C.L., 95\% C.L., and 99.7\% C.L. between each other for the parameter space of model 1.}
\label{fig:6009}
\end{center}
\end{figure}

\begin{figure}
\begin{center}
  \includegraphics[scale=0.35]{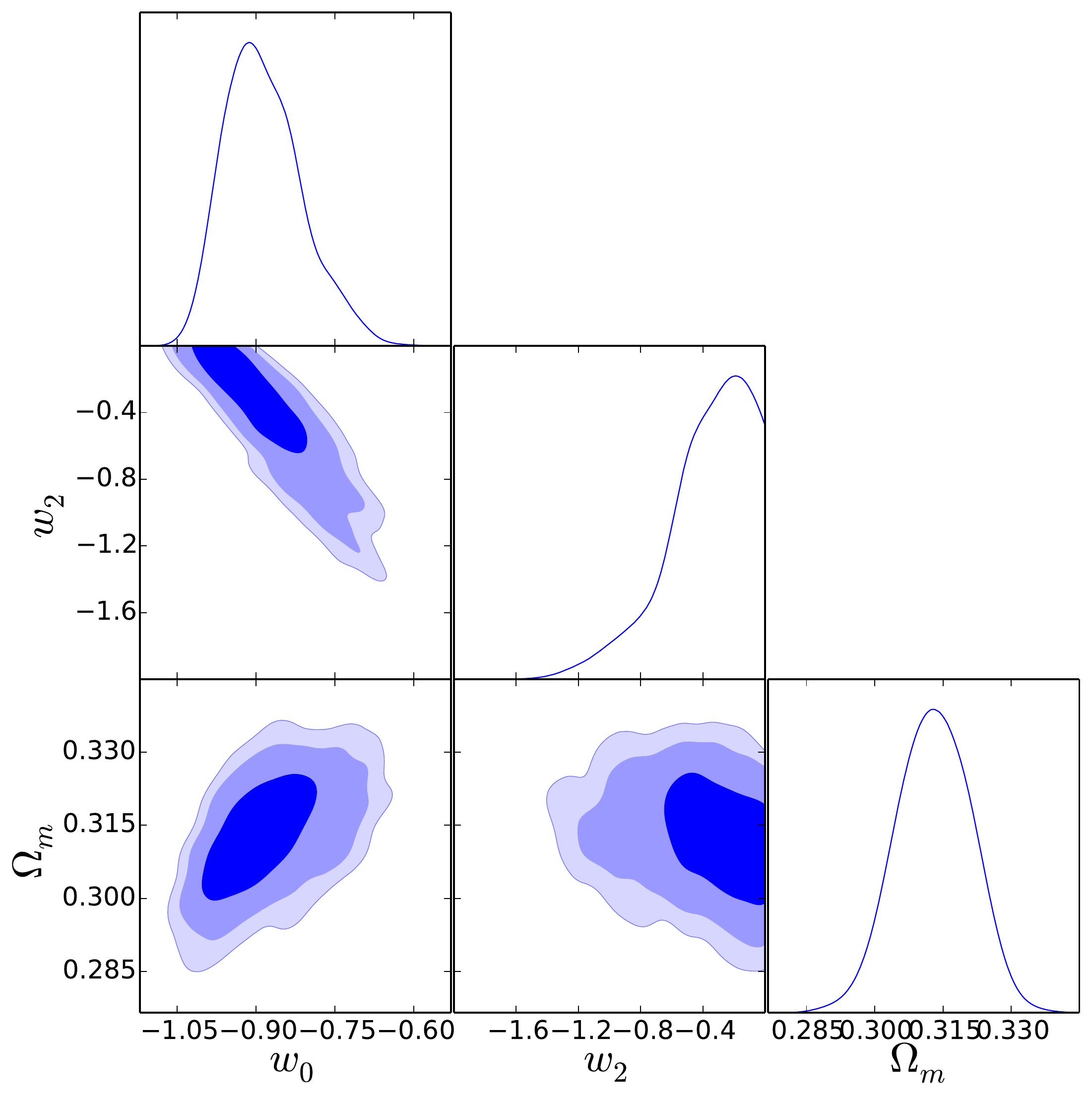}
\caption{1D marginalized distributions on individual parameters and 2D contours with 68\% C.L., 95\% C.L., and 99.7\% C.L. between each other for the parameter space of model 2.}
\label{fig:6010}
\end{center}
\end{figure}

\begin{figure}
\begin{center}
 \includegraphics[scale=0.35]{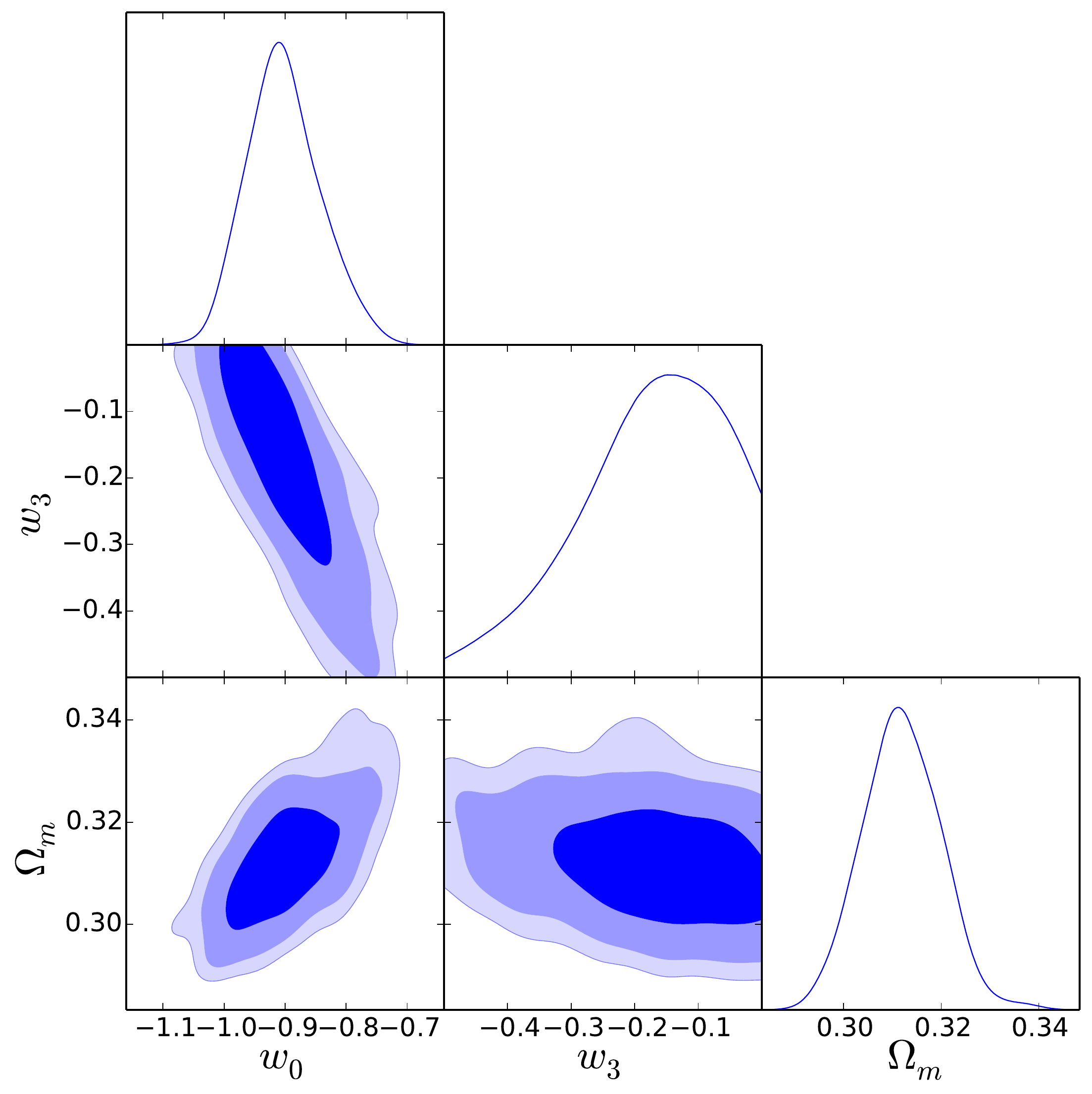}
\caption{1D marginalized distributions on individual parameters and 2D contours with 68\% C.L., 95\% C.L., and 99.7\% C.L. between each other for the parameter space of model 3.}
\label{fig:6008}
\end{center}
\end{figure}

After running eight chains in parallel on computer, the mean values with errors and best-fit values, as well as the logarithmic Bayesian evidence, are presented in TABLE \ref{tab:table1}, \ref{tab:table2}, \ref{tab:table3}, for three models respectively. One dimensional (1D) marginalized distributions of parameters and 2D contours with 68\% C.L., 95\% C.L., and 99.7\% C.L. are presented in FIG. \ref{fig:6009}, \ref{fig:6010}, \ref{fig:6008}.

We note that the constraint of $w_i$ are not so compact comparing to the other parameters, $\Omega_{m0}$, $w_0$, so is the case in Wang and Meng's work \cite{ref:3}. We consider it the intrinsic properties of the time-dependent models, while those models are just phenomenological modification of the $\Lambda$CDM scenario or approximate description of the observation.

\begingroup
\begin{center}
\begin{table}
\renewcommand{\arraystretch}{1.8}
\begin{tabular}{ccc}
\hline\hline
Parameters & Mean with errors & Best fit \\ \hline
$\Omega_{m0}$ & $0.312_{-0.00868-0.0167-0.0195}^{+0.00850+0.0159+0.0242}$ & $0.312$\\
$w_0$ & $-0.893_{-0.0899-0.137-0.153}^{+0.0647+0.152+0.202}$ & $-0.941$\\
$w_1$ & $0.300_{-0.260-0.300-0.300}^{+0.109+0.355+0.536}$ & $0.0916$\\
\hline
$\log \mathcal{Z}_1$ & $-1675.4192 \pm 0.3444$ & $ $\\
\hline\hline
\end{tabular}
\caption{The mean values with $1\sigma$, $2\sigma$, $3\sigma$ error bars and best-fit values for the parameter space of model 1. The last column is its logarithmic Bayesian evidence.}\label{tab:table1}
\end{table}
\end{center}
\endgroup

\begingroup
\begin{center}
\begin{table}
\renewcommand{\arraystretch}{1.8}
\begin{tabular}{ccc}
\hline\hline
Parameters & Mean with errors & Best fit \\ \hline
$\Omega_{m0}$ & $0.313_{-0.00835-0.0163-0.0228}^{+0.00848+0.0156+0.0205}$ & $0.310$\\
$w_0$ & $-0.885_{-0.0865-0.134-0.151}^{+0.0594+0.154+0.210}$ & $-0.909$\\
$w_2$ & $-0.383_{-0.0967-0.546-0.826}^{+0.377+0.383+0.383}$ & $-0.336$\\
\hline
$\log \mathcal{Z}_2$ & $-1674.6744 \pm 0.3430$ & $ $\\
\hline\hline
\end{tabular}
\caption{The mean values with $1\sigma$, $2\sigma$, $3\sigma$ error bars and best-fit values for the parameter space of model 2. The last column is its logarithmic Bayesian evidence.}\label{tab:table2}
\end{table}
\end{center}
\endgroup

\begingroup
\begin{center}
\begin{table}
\renewcommand{\arraystretch}{1.8}
\begin{tabular}{ccc}
\hline\hline
Parameters & Mean with errors & Best fit \\ \hline
$\Omega_{m0}$ & $0.311_{-0.00821-0.0157-0.0192}^{+0.00812+0.0159+0.0251}$ & $0.310$\\
$w_0$ & $-0.901_{-0.0683-0.115-0.137}^{+0.0549+0.124+0.163}$ & $-0.880$\\
$w_3$ & $-0.179_{-0.0675-0.209-0.320}^{+0.152+0.179+0.179}$ & $-0.255$\\
\hline
$\log \mathcal{Z}_3$ & $-1673.6427 \pm 0.3399$ & $ $\\
\hline\hline
\end{tabular}
\caption{The mean values with $1\sigma$, $2\sigma$, $3\sigma$ error bars and best-fit values for the parameter space of model 3. The last column is its logarithmic Bayesian evidence.}\label{tab:table3}
\end{table}
\end{center}
\endgroup

The CosmoChord also generates covariance matrixes for the vector of random variables. The covariance coefficients of $(\Omega_{m0},w_0,w_i)$ are displayed in matrix form (\ref{3-3}), (\ref{3-4}), (\ref{3-5}) for three models respectively, which are all on $1\sigma$ confidence level.

\begin{eqnarray}\label{3-3}
Cov_1 = \left(
\begin{array}{ccc}
 6.670\times 10^{-6} & -2.842\times 10^{-5} & 1.348\times 10^{-4} \\
 -2.842\times 10^{-5} & 5.714\times 10^{-3} & 1.184\times 10^{-2} \\
 1.348\times 10^{-4} & 1.184\times 10^{-2} & 3.690\times 10^{-2}
\end{array}
\right), \nonumber \\
\end{eqnarray}

\begin{equation}\label{3-4}
Cov_2 =\left(
\begin{array}{ccc}
 7.593\times 10^{-6} & 2.249\times 10^{-5} & -3.229\times 10^{-4} \\
 2.249\times 10^{-5} & 5.551\times 10^{-3} & -1.754\times 10^{-2} \\
 -3.229\times 10^{-4} & -1.754\times 10^{-2} & 7.462\times 10^{-2}
\end{array}
\right),
\end{equation}

\begin{equation}\label{3-5}
Cov_3 =\left(
\begin{array}{ccc}
 7.410\times 10^{-6} & -1.665\times 10^{-5} & -1.048\times 10^{-4} \\
 -1.665\times 10^{-5} & 3.736\times 10^{-3} & -5.253\times 10^{-3} \\
 -1.048\times 10^{-4} & -5.253\times 10^{-3} & 1.229\times 10^{-2}
\end{array}
\right).
\end{equation}

\section{Analysis on models} \label{sec:analysis}

\begin{figure}
\begin{center}
 \includegraphics[width=8cm,height=6cm]{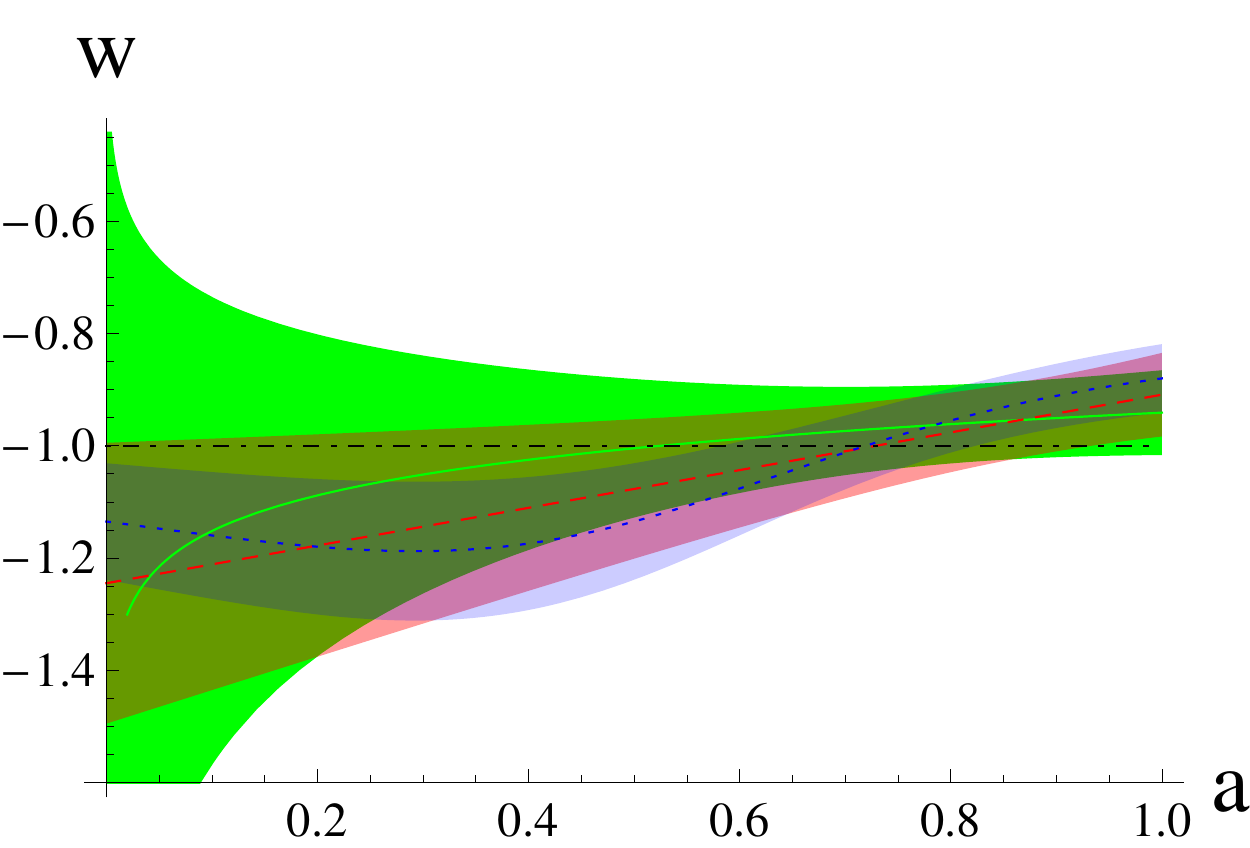}
\caption{\label{fig:wa} The evolving curve of EOS with $1\sigma$ error range. Green-solid, red-dashed, blue-dotted, black-dot-dashed line correspond to model 1, model 2, model 3 and $\Lambda$CDM respectively. The values of $a$ range from $0$ to $1$.}
\end{center}
\end{figure}

Before inspecting the power of diagnostics stated in Section \ref{sec:models}, the evolving curve of EOS with $1\sigma$ error range for three models are plotted in FIG. \ref{fig:wa}.
Just from the best-fit curves, it seems that all three EOSs share a same trend: less than $-1$ at early stage, and go across it when $a=0.5\sim0.7$.
However, model 1 is extremely uncertain at early stage in comparison to the other two when we consider errors.
The region of model 3 is so compact during all the evolving history that we can preliminary judge that model 3 is an optimal one in this set.
For further analysis, substituting the best-fit values of the random variables $(\Omega_{m0},w_0,w_i)$ in TABLE \ref{tab:table1}, \ref{tab:table2}, \ref{tab:table3}, and the covariance matrices (\ref{3-3}), (\ref{3-4}), (\ref{3-5}) into $S^{(1)}_3$ (\ref{2-16}), $S^{(1)}_4$ (\ref{2-18}), $S^{(1)}_5$ (\ref{2-20}), $S^{(2)}_5$ (\ref{2-21}), then we obtained the modified phase diagrams.
The original phase diagrams without error bars are also displayed as a comparison to see improvements on discriminating different dark energy models.

\begin{figure}
\begin{center}
 \includegraphics[width=8cm,height=6cm]{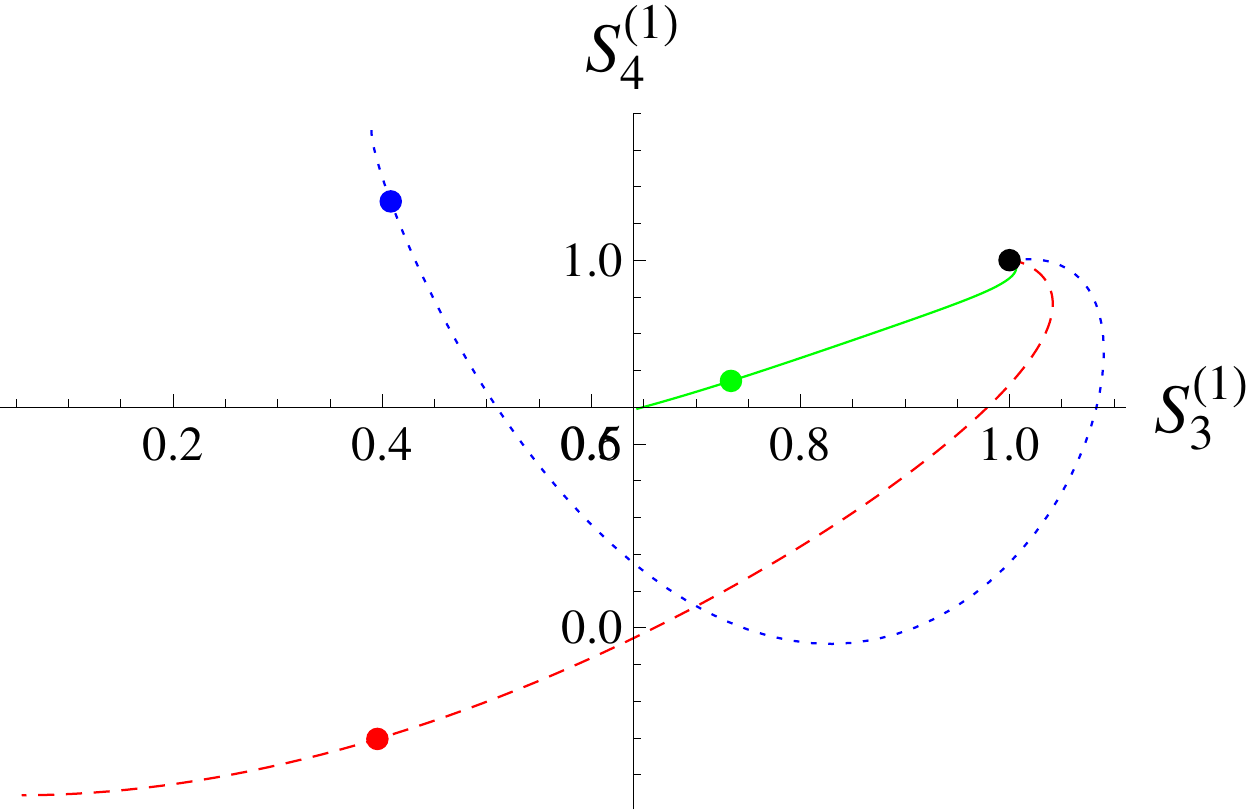}
\caption{\label{fig:rs41} The phase diagram of $\{S^{(1)}_3, S^{(1)}_4\}$ without error bars. Green-solid, red-dashed, blue-dotted line correspond to model 1, model 2, model 3 and the points label the present situations respectively. Specifically, the black point denotes the $\Lambda$CDM model. The values of $a$ range from $0$ to $1.2$.}
\end{center}
\end{figure}

\begin{figure}
\begin{center}
 \includegraphics[width=8cm,height=6cm]{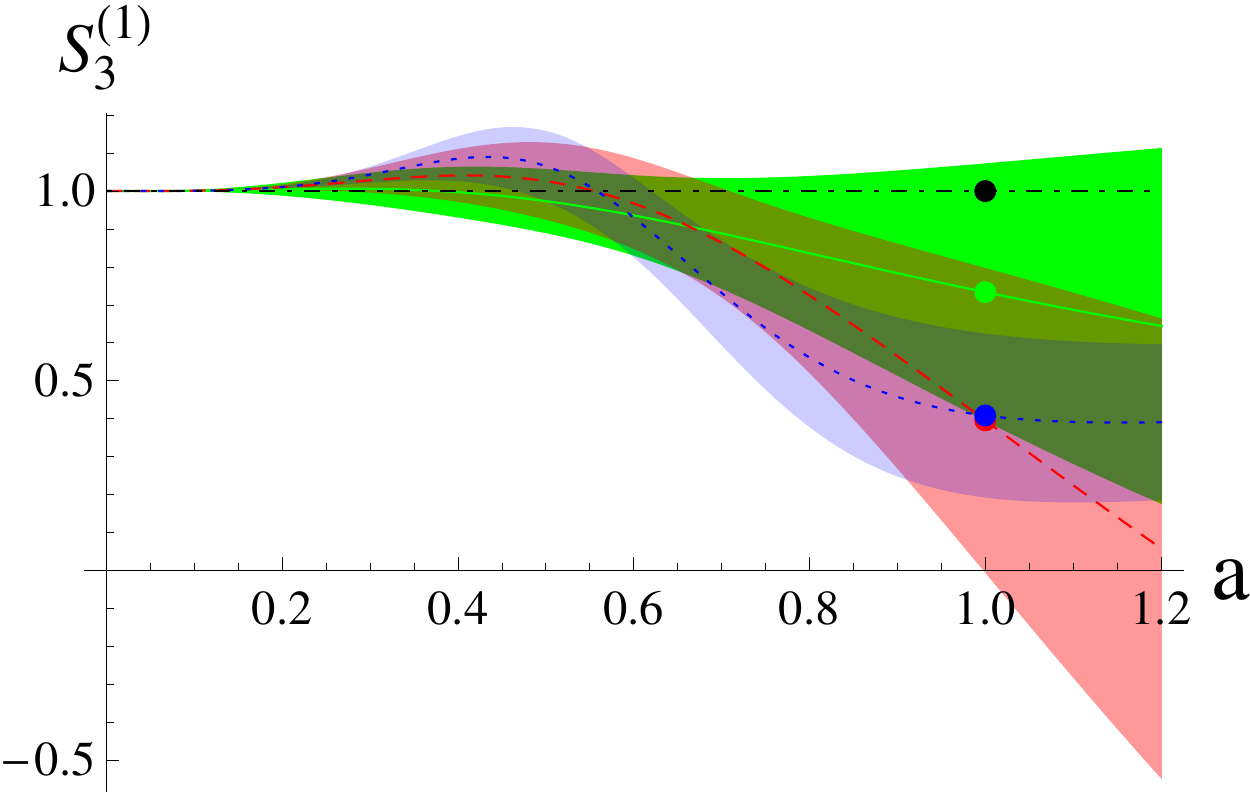}
\caption{\label{fig:ra} The evolving diagram of $S^{(1)}_3$ with $1\sigma$ error range. Green-solid, red-dashed, blue-dotted, black-dot-dashed line correspond to model 1, model 2, model 3 and $\Lambda$CDM respectively. The values of $a$ range from $0$ to $1.2$.}
\end{center}
\end{figure}

\begin{figure}
\begin{center}
 \includegraphics[width=8cm,height=6cm]{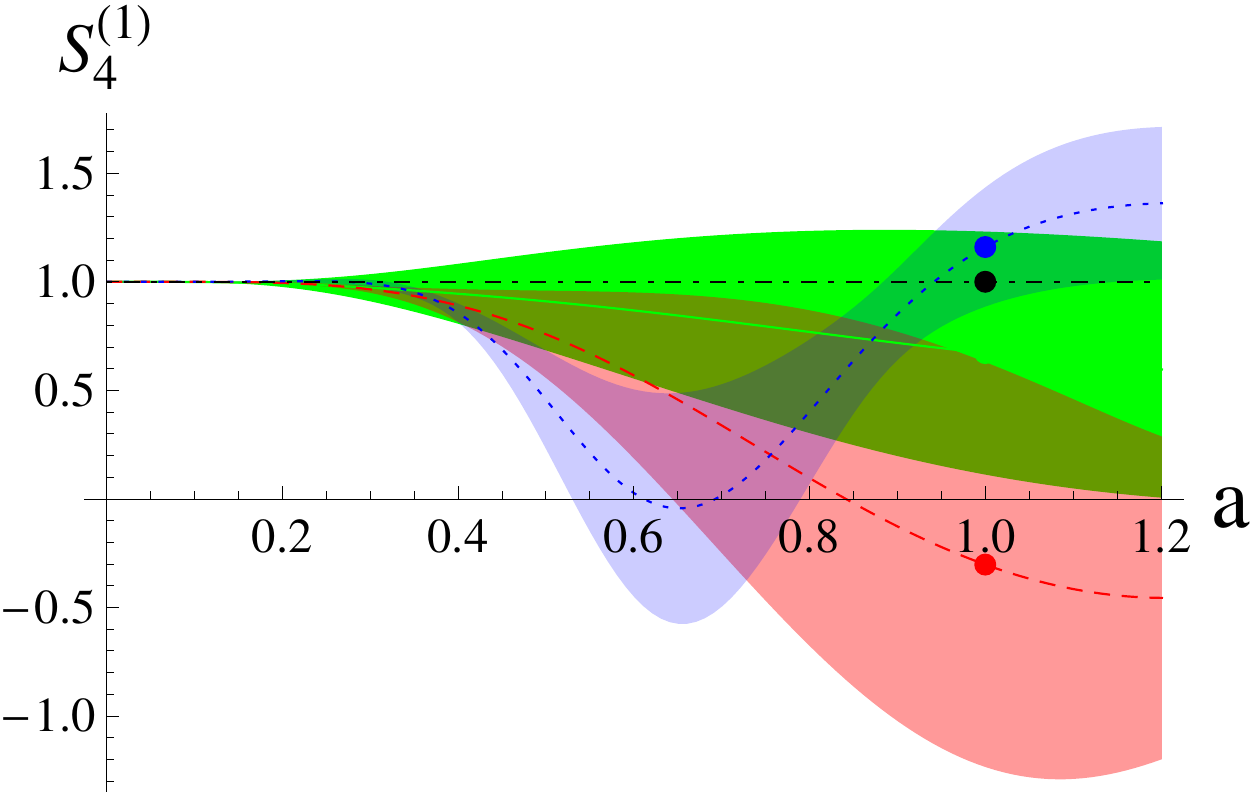}
\caption{\label{fig:s41} The evolving diagram of $S^{(1)}_4$ with $1\sigma$ error range. Green-solid, red-dashed, blue-dotted, black-dot-dashed line correspond to model 1, model 2, model 3 and $\Lambda$CDM respectively. The values of $a$ range from $0$ to $1.2$.}
\end{center}
\end{figure}

The phase diagram of $\{S^{(1)}_3, S^{(1)}_4\}$ without error bars for three models is showed in FIG. \ref{fig:rs41}.
It is obvious that all the three models together with the $\Lambda$CDM can be well distinguished from each other at the present period.
FIG. \ref{fig:ra} is the evolution of $S^{(1)}_3$ in terms of $a$ with $1\sigma$ error bars.
In this figure, it is hard to discriminate anyone from the other two, because their error regions overlap each other.
But model 1 and model 2 show considerable uncertainty at present and future period, while the $1\sigma$ region of model 3 is some relatively compact.
This is coincident with the conclusion from FIG. \ref{fig:wa}.
In addition, model 3 shows slight inflection at about $a\sim5$ and $a\sim 1$.
This feature can be seen from FIG. \ref{fig:s41} more obviously.
When compare to the fiducial model, it is easy to distinguish model 2 from the $\Lambda$CDM roughly at $a=0.7$, and the deviation tends to increase with time.
This is not a good property because the $\Lambda$CDM cosmology fits the current observational data sets very well even it has some issues. From these features, it is not easy to distinguish each dark energy model from the other models clearly.

\begin{figure}
\begin{center}
 \includegraphics[width=8cm,height=6cm]{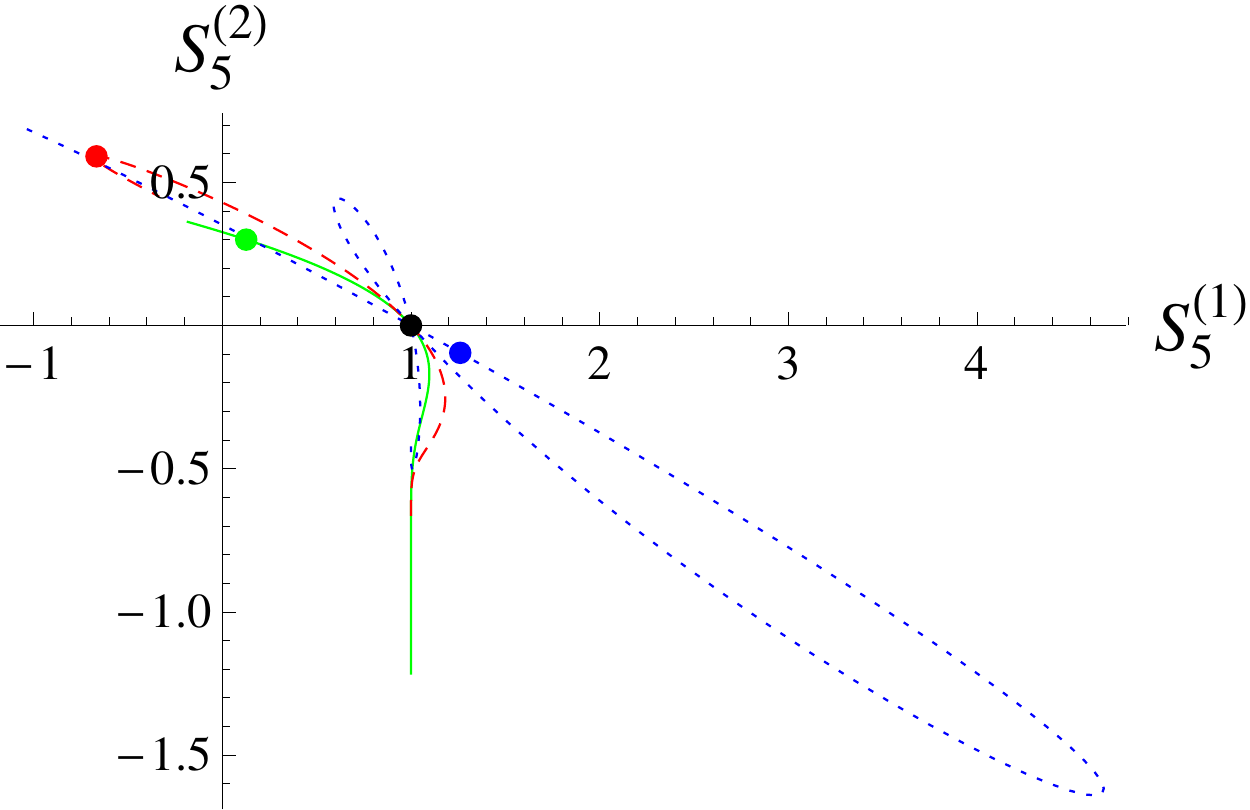}
\caption{\label{fig:s51s52} The phase diagram of $\{S^{(1)}_5, S^{(2)}_5\}$ without error bars. Green-solid, red-dashed, blue-dotted line correspond to model 1, model 2, model 3 and the points label the present situations respectively. Specifically, the black point denotes the $\Lambda$CDM model. The values of $a$ range from $0$ to $1.2$.}
\end{center}
\end{figure}

\begin{figure}
\begin{center}
 \includegraphics[width=8cm,height=6cm]{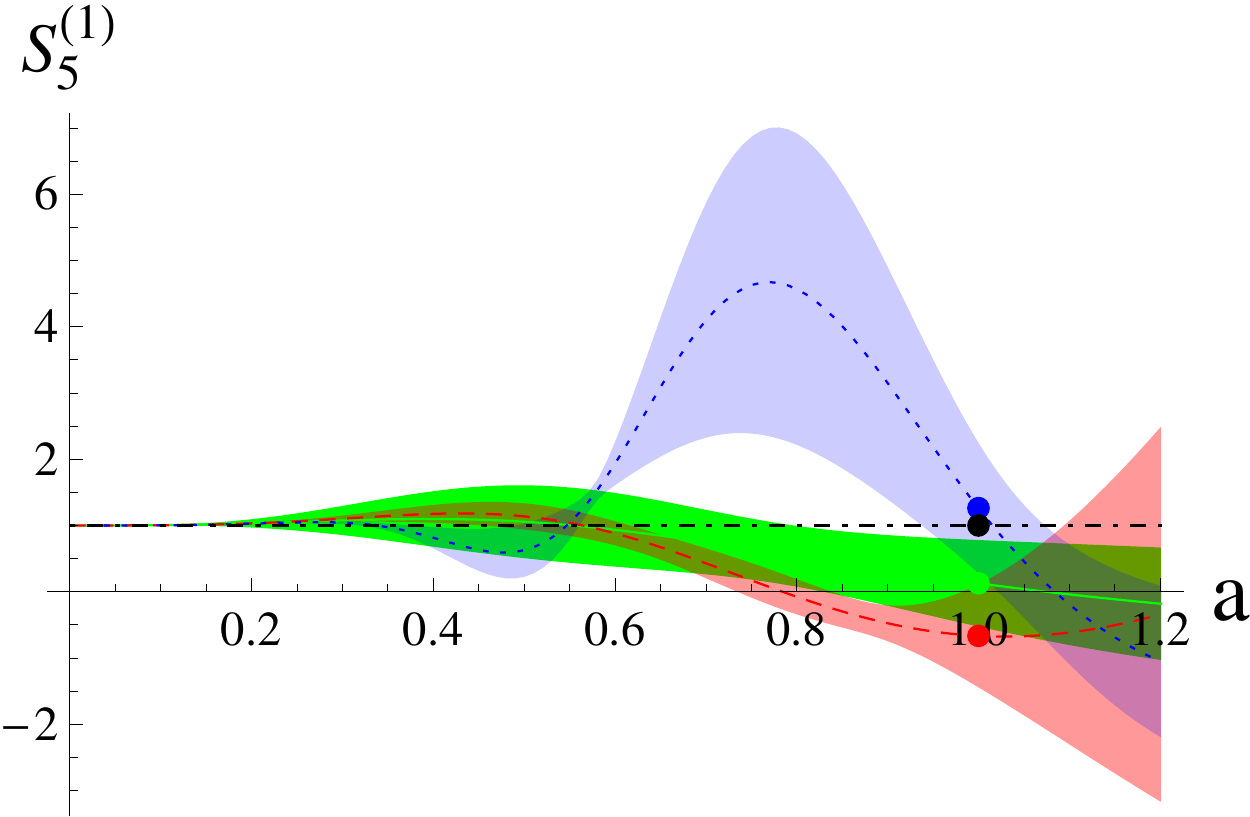}
\caption{\label{fig:s51} The evolving diagram of $S^{(1)}_5$ with $1\sigma$ error range. Green-solid, red-dashed, blue-dotted, black-dot-dashed line correspond to model 1, model 2, model 3 and $\Lambda$CDM respectively. The values of $a$ range from $0$ to $1.2$.}
\end{center}
\end{figure}

\begin{figure}
\begin{center}
 \includegraphics[width=8cm,height=6cm]{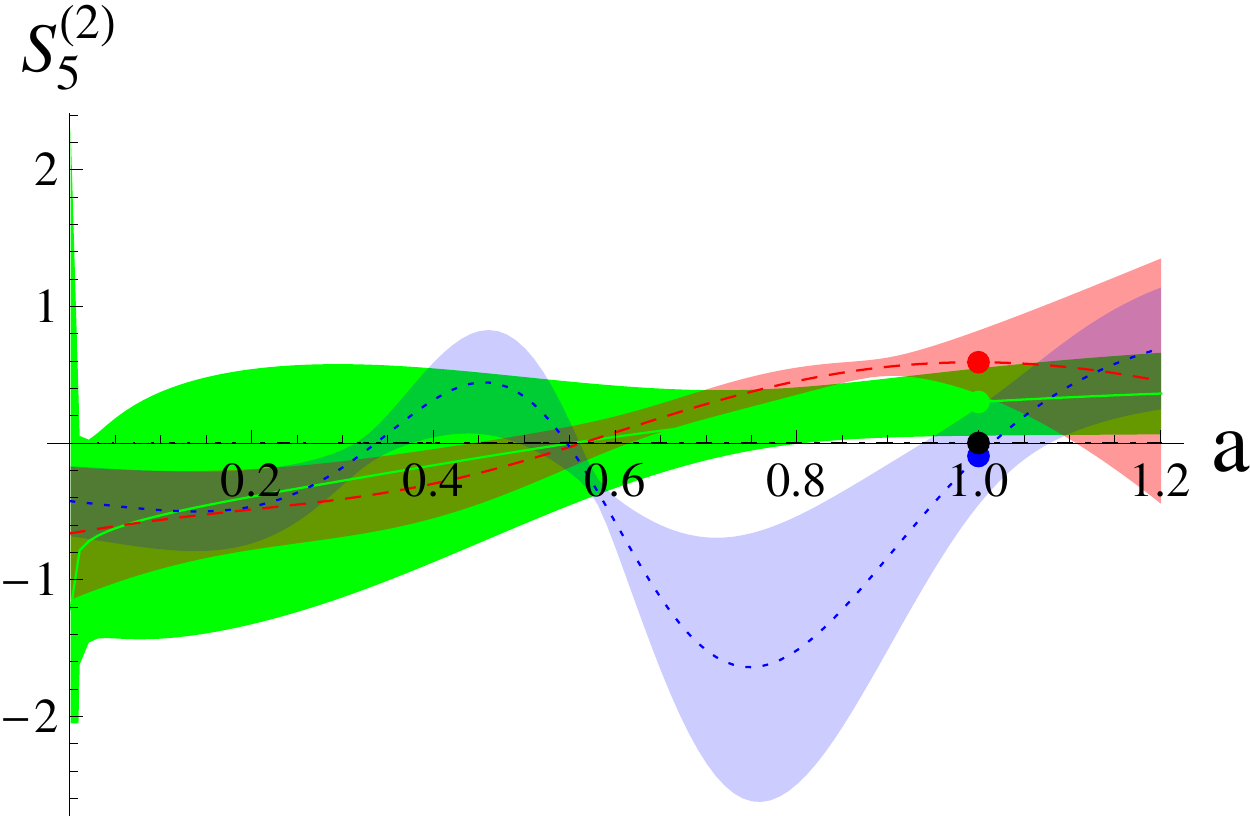}
\caption{\label{fig:s52} The evolving diagram of $S^{(2)}_5$ with $1\sigma$ error range. Green-solid, red-dashed, blue-dotted, black-dot-dashed line correspond to model 1, model 2, model 3 and $\Lambda$CDM respectively. The values of $a$ range from $0$ to $1.2$.}
\end{center}
\end{figure}

Subsequently, we investigate another pair of statefinder, $\{S^{(1)}_5, S^{(2)}_5\}$.
Their phase diagrams without error bars are plotted in FIG. \ref{fig:s51s52}.
It is easy to see that the blue point quite closes to the black point, which means that model 3 is a potential alternative to the $\Lambda$CDM model at present period.
The trajectory of model 3 is completely different from the other two.
It is worth noting that model 3 goes through the fixed point (1, 0) which corresponds to the $\Lambda$CDM scenario more than one time.
In FIG. \ref{fig:s51}, model 3 shows significant fluctuation, and it is easy to distinguish it from the remaining models at $a=0.6\sim1$ even with $1\sigma$ error range while model 1 presents a stable and compact feature. However, it is hasty to say model 1 is better than model 3, because model 1 is singular at early stage in FIG. \ref{fig:s52}.
Conversely, model 2 is more uncertainty in the future.

The above analysis from statefinder with error bars seems to support model 3 as the best one during all the evolving history from $a=0$ to $a=1.2$.
Immediately the Bayesian evidence prove this point of view in a powerfully numerical way.
First, it is easy to see that $\log \mathcal{Z}_3 > \log \mathcal{Z}_2 > \log \mathcal{Z}_1$, then we calculate the difference of $\log \mathcal{Z}$ for three models:
\begin{eqnarray}
  \log \mathcal{Z}_3-\log \mathcal{Z}_2 = 1.0317,\label{4-1} \\
  \log \mathcal{Z}_3-\log \mathcal{Z}_1 = 1.7765,\label{4-2} \\
  \log \mathcal{Z}_2-\log \mathcal{Z}_1 = 0.7448.\label{4-3}
\end{eqnarray}
According to Harold Jeffrey's scale (ref) in TABLE \ref{tab:jeffrey}, model 3 is significantly better than model 1 and model 2, while model 1 and model 2 are inconclusive. This result coincides with the analysis from statefinder.

\section{Conclusion}  \label{sec:conclusion}

Since there is so many dark energy models, one of the most interesting work is model selection, that is, distinguishing them from each other and choosing the most reasonable or better model based on the current observational data sets.
In this work, three TDDE models are introduced, and then we combined two completely different approaches to distinguish them.
First, we improved the statefinder by adding error bars via the error propagation equation.
Then, using the cosmic observational data sets: $Planck$ 2015 low-$l$, TT, EE, TE, lensing, joint SNIa, BAO, OHD in CosmoChord, the three TDDE models's parameters are constrained.
Meanwhile, the Bayesian evidence for different models are also given.

Although the constraint of $w_i$ are not so compact comparing to the other parameters, $\Omega_{m0}$, $w_0$, using the modified figures of $S^{(1)}_3, S^{(1)}_4$, $S^{(1)}_5, S^{(2)}_5$, we could tell that the accuracy of $w_i$ is enough to obtain a meaningful consequence.
When the error bars are not included, different models have different trajectories, the TDDE models can be well distinguished from each other and the $\Lambda$CDM model. However, when the error bars are taking into consideration, the trajectories are changed to error region, and different model's error region have some intersection. One can compare the trend of evolution as well as the compactness of the error distribution region of each model in different period of time, which reflect the closeness of fit between the model and the observational data. Therefore, it maybe unworkable to distinguish different models without consider the error bars.

After carefully analysing, we find that model 3 is an optimal one, because the error region of it is relativity compact during all the evolving history.
The other two models usually have a significant uncertainly either at early stage or in the future.
This conclusion is also supported by the results of the Bayesian evidence.
According to Harold Jeffrey's scale, model 3 is significantly better than model 1 and model 2, while model 1 and model 2 are inconclusive.
All of this encourages us to use model 3 as a preferential candidate in dark energy investigation rather than the other two.

\acknowledgements

We thank an anonymous referee for helpful improvement of this paper. L. Xu's work is supported in part by NSFC under the Grants No. 11275035.

\end{document}